\documentclass[aps,pra,superscriptaddress,showpacs,showkeys,amsfonts,amssymb,amsmath]{revtex4}
\usepackage{amsmath,amssymb}
\usepackage{epsfig}
\newcommand{\n}{\nonumber}
\newcommand{\be}{\begin{eqnarray}}
\newcommand{\ee}{\end{eqnarray}}

\begin{document}

\title{Extensions of  a class of similarity solutions of Fokker-Planck equation with time-dependent coefficients and fixed/moving boundaries}

\author{Choon-Lin Ho}
\affiliation{Department of Physics, Tamkang University,
 Tamsui 25137, Taiwan, R.O.C.}

\author{Ryu Sasaki\footnote{Present affiliation \
 Department of Physics, Shinshu University,
     Matsumoto 390-8621, Japan}}
\affiliation{
  Center for Theoretical Sciences,
    National Taiwan University, Taipei, Taiwan, R.O.C.}

\begin{abstract}

A general formula in closed form to obtain exact similarity solutions  of the Fokker-Planck
equation with both time-dependent drift and diffusion coefficients was recently presented by Lin and Ho [ Ann. Phys. \textbf{327}, 386 (2012);
J. Math. Phys.  \textbf{54}, 041501 (2013)].  In this paper we extend the class of exact solutions by exploiting certain properties of the general formula.
\end{abstract}




\maketitle


\section{Introduction}

One of the basic tools which
are widely used for studying the 
effects of fluctuations in
macroscopic systems is the Fokker-Planck equation (FPE) \cite{RIS:1996}.
This equation has found applications in such diverse areas as
physics, chemistry, hydrology, biology, finance and
others. Because of its broad applicability, it is therefore of
great interest to obtain solutions of the FPE for various physical
situations.

Generally, it is not easy to find analytic solutions of the FPE, except 
for a few simple cases,
such as linear drift and constant diffusion coefficients. In most cases, one can only solve the equation
approximately, or numerically.  Most of these methods, however, are concerned
only with FPEs with time-independent diffusion and drift
coefficients (for a review of these methods, see eg. Ref. ~\cite{RIS:1996}).

Solving the FPEs with time-dependent drift and/or diffusion
coefficients is in general an even more difficult task. It is
therefore not surprising that the number of papers on such kind of
FPE is far less than that on the FPE with time-independent
coefficients. 

Recently, we have presented a general formula in closed form for a class of exact solutions of the FPEs based on the 
similarity method, 
both for fixed and moving boundaries \cite{LH,Ho}. 
One advantage of the similarity method is that it allows one to
reduce the FPE to an
ordinary differential equation which is generally easier to solve, provided that the FPE
possesses proper scaling property under certain scaling
transformation of the basic variables.   It is
interesting to find, by the natural requirement that the
probability current density vanishes at the boundary, that the
resulting ordinary differential equation is exactly integrable, and the
probability density function can be given in closed form.  Our work has extended the number of exact solutions of FPEs with
time-dependent drift and/or diffusion
coefficients. 

In this note, we show that the class of exact solutions of the FPEs can be further extended by exploiting 
certain properties of the general formula presented in \cite{LH,Ho}.

\section{Scaling of Fokker-Planck equation}

We first review  the scaling form of the FPE \cite{LH,Ho}.
The general form of the FPE in $(1+1)$-dimension is
\begin{eqnarray}\label{E2.1}
\frac{\partial W(x,t)}{\partial t}=\Big[-\frac{\partial}{\partial x}
D^{(1)}(x,t)+\frac{\partial^2}{\partial x^2}D^{(2)}(x,t)\Big]W(x,t)\;,
\end{eqnarray}
where $W(x,t)$ is the probability distribution function (PDF),
$D^{(1)}(x,t)$ is the drift coefficient and $D^{(2)}(x,t)$ the
diffusion coefficient. The drift coefficient represents the
external force acting on the particle, while the diffusion
coefficient accounts for the effect of fluctuation. $W(x,t)$ as a
probability distribution function should be normalized, $i.e.$,
$\int_{\textstyle\mbox{\small{domain}}}W(x,t)\,dx=1$ for $t\geq
0$.

We shall be interested in seeking similarity solutions of the FPE, which are possible if
the FPE is invariant under  the scale transformation
\begin{align}\label{E2.2}
&\bar{x}=\varepsilon^a x\;\;\;,\;\;\;\bar{t}=\varepsilon^b t,
\end{align}
where $\varepsilon
>0$, $a$ and $b$ are real parameters. Suppose
under this transformation, the PDF and
the two coefficients scale as
\begin{align}\label{E2.3}
\bar{W}(\bar{x},\bar{t})=\varepsilon^c W(x,t),~
\bar{D}^{(1)}(\bar{x},\bar{t})=\varepsilon^d
D^{(1)}(x,t), ~\bar{D}^{(2)}(\bar{x},\bar{t})=\varepsilon^e
D^{(2)}(x,t).
\end{align}
Here $c$, $d$ and $e$ are also some real parameters.  It can be checked that
the transformed equation in terms of the new variables has the same functional form as Eq.(\ref{E2.1})
if the scaling indices satisfy $b=a-d=2a-e$. In this case, the second order FPE can  be transformed into an
ordinary differential equation which is easier to solve. Such reduction is 
achieved
through a new independent variable $z$ (called 
a similarity variable), which is
certain combination of the old independent variables such that
it is scaling invariant, i.e., no appearance of 
the parameter
$\varepsilon$, as a scaling transformation is performed. Here
the similarity variable $z$ is defined by
\begin{align}\label{E3.1}
z\equiv\frac{x}{t^{\alpha}}\;,\;\;\;\mbox{where}\;\;\;
\alpha=\frac{a}{b}\;\;\;\mbox{and}\;\;\;a\;,b\neq 0\;.
\end{align}
For $a\;,b\neq 0$, one has $\alpha\neq 0\;,\infty$.

The scaling form of the PDF
can be taken as \cite{LH}
\begin{align}\label{E3.2}
W(x,t)=t^{\alpha\frac{c}{a}}y(z)\;,
\end{align}
where $y(z)$ is a function of $z$. The normalization of the distribution function is
\begin{align}
\int_{\mbox{\small{domain}}}\,W(x,t)\,dx=\int_{\mbox{\small{domain}}}\,
\Big[t^{\alpha(1+\frac{c}{a})}\,y(z)\Big]\,dz=1\;.
\label{norm}
\end{align}
For the above relation to hold at all $t\geq 0$, the power of $t$
should vanish, and so one must have $c=-a$, and thus
\begin{equation}
W(x,t)=t^{-\alpha}y(z). \label{W-scale}
\end{equation}
Similar consideration leads to the following scaling forms of the
drift  and diffusion coefficients
\begin{align}\label{E3.3}
D^{(1)}(x,t)=t^{\alpha-1}\rho_1(z)\;\;\;,\;\;\;D^{(2)}(x,t)=t^{2\alpha-1}\rho_2(z)\;,
\end{align}
where $\rho_1(z)$ and $\rho_2(z)$ are scale invariant functions of $z$.

With Eqs.~(\ref{E3.1}), (\ref{W-scale}) and (\ref{E3.3}), the FPE is
reduced to
\begin{align}\label{E3.4}
\rho_2(z)\,y''(z)+\Big[2\rho_2'(z)-\rho_1(z)+\alpha z\Big]\,y'(z)
+\Big[\rho_2''(z)-\rho_1'(z)+\alpha \Big]\,y(z)=0\;,
\end{align}
where the prime denotes the derivative with respect to $z$.
It is really interesting to realize  that Eq.~(\ref{E3.4}) is exactly integrable.
Integrating it once, we get
\begin{equation}
\rho_2(z) y^\prime(z) + \left[\rho_2^\prime (z) - \rho_1 (z)+
\alpha z\right] y(z)=C, \label{y1}
\end{equation}
where $C$ is an integration constant.
Solution of Eq.~(\ref{y1}) is
\begin{eqnarray}
y(z)&=&\left(C^\prime+C\int^z dz \frac{e^{-\int^z\,dz f(z)}}{\rho_2(z)}\right)\,\exp\left(\int^zdz f(z)\right),\nonumber\\
f(z)&\equiv& \frac{\rho_1(z)-\rho_2^\prime(z)-\alpha
z}{\rho_2(z)},~~~\rho_2(z)\neq 0,
\label{y-soln}
\end{eqnarray}
where $C^\prime$ is an integration constant.

We shall consider boundaries which are impenetrable.  At such boundaries,
the probability density and the associated probability
current density must vanish.
This in turn implies that $C=0$,  and the PDF $W(x,t)$ is given by
\begin{eqnarray}
W(x,t) &=&
 t^{-\alpha} y(z),\n\\
y(z)&\equiv& 
A\exp\left(\int^z
dz\,f(z)\right)_{z=\frac{x}{t^\alpha}},\label{W-soln}
\end{eqnarray}
where $A$ is the normalization constant.  It is
interesting to see that the similarity solution of the FPE can be
given in such an analytically 
closed form.  Exact similarity solutions
of the FPE can be obtained as long as $\rho_1(z)$ and $\rho_2(z)$
are such that the function $f(z)$ in Eq.~(\ref{W-soln}) is an
integrable function and the resulted $W(x,t)$ is normalizable.
Equivalently, for any integrable function $f(z)$ such that
$W(x,t)$ is normalizable, if one can find a function $\rho_2(z)$
($\rho_1(z)$ is then determined by $f(z)$ and $\rho_2(z)$), then
one obtains an exactly solvable FPE with similarity solution given
by Eq.~(\ref{W-soln}).  

Some interesting cases of such FPE on the real line $x\in (-\infty,\infty)$ and
the half lines $x\in [0,\infty)$ and $x\in (-\infty,0]$  were discussed in Ref.~\cite{LH}.
These domains admit similarity solutions because their boundary points are the fixed points of the scaling transformation considered.
This indicates that similarity solutions are not possible for other finite domains,
unless  its boundary points scale accordingly. This leads to FPE with moving boundaries.
Examples of such FPEs are presented in Ref.~\cite{Ho}.

%
%

\section{Extension of known classes}

From Eq.~(\ref{W-soln}) it is evident that for any solvable solution given by $\rho_1(z)$ and $\rho_2(z)$, 
there exist other solvable systems with a positive function $Q(z)$
\begin{eqnarray}
\tilde{\rho}_1(z) &\equiv& \rho_1(z) +  \rho_2(z)\,\frac{d}{dz}\ln Q(z),\n\\
\tilde{\rho}_2(z)&= &\rho_2(z),
\end{eqnarray}
and
\begin{equation}
\tilde{f}(z)=f(z)+\frac{d}{dz}\ln Q(z)
\end{equation}
The corresponding probability density is
\begin{equation}
\tilde{W}(x,t)=\tilde{A}Q(z)\, t^{-\alpha} y(z),
\end{equation}
where $\tilde{A}$ is a new normalization constant.
Hence a new extended system can be obtained from the original system by choosing an appropriate function 
$Q(z)$ as long as $\tilde{W}(x,t)$ is normalizable. This can be done by, say, 
looking up appropriate integrable definite integrals in \cite{GR}.

The above discussion gives a general recipe of finding new exactly FPEs.  
 To illustrate the idea, let 
us take for example the case presented in Sect.~4 of \cite{LH}, with  $\rho_1(z)$ 
and $\rho_2(z)$ given by (with some changes of the notations of the coefficients)
\begin{align}\label{E4.2}
\rho_1(z)=\lambda z-\mu\;\;\;,\;\;\;\rho_2(z)=\sigma\;,
\end{align}
where $\lambda$, $\mu$ and $\sigma$ are real constants. This choice
of $\rho_1 (z)$ and $\rho_2 (z)$ generates 
 the following drift and
diffusion coefficients:
\begin{align}\label{E4.3}
D^{(1)}(x,t)=\lambda\,\frac{x}{t}-\mu\,t^{\alpha-1}\;\;\;,\;\;\;
D^{(2)}(x,t)=\sigma\,t^{2\alpha-1}\;.
\end{align}
From Eq.~(\ref{W-soln}), the function $y(z)$ is
\begin{eqnarray}
y(z) &\propto& \left\{
\begin{array}{ll}
\mbox{exp}\Big\{\frac{1}{\sigma}\left[\left(\lambda-\alpha\right)\frac{z^2}{2}-\mu z\right]\Big\}, 
&\lambda\neq \alpha;\\
& \\
\mbox{exp}\Big\{-\frac{\mu}{\sigma}z\Big\},  &\lambda=\alpha.
\end{array}
\right.
\label{case-I}
\end{eqnarray}

We shall discuss these two cases separately.

\subsection{Examples with $\lambda=\alpha$}

In this case, the normalized solution is
\begin{align}
W(x,t)=\Big|\frac{\mu}{\sigma\,t^{\alpha}}\Big|\,\exp\left(-\frac{\mu}{\sigma\,t^{\alpha}}\,x\right),
\label{W1}
\end{align}
where it is valid in $x\geq 0$ for $(\mu/\sigma)>0$; $x\leq 0$ for $(\mu/\sigma)<0$.

For simplicity and clarity of presentation, let us take $\mu>0,~\sigma=1$, i.e.,   
$\rho_2(z)=1$.  
A new extended system
can be obtained by taking, 
for example,
\begin{equation}
Q(z)=z^{\nu-1},~~\nu>1.
\end{equation}
By using the identity
\begin{equation}
\int^\infty_{0}\, x^{\nu-1}\,e^{-\mu x}\,dx=\frac{\Gamma(\nu)}{\mu^\nu},
\end{equation}
we obtain the corresponding normalized PDF as
\begin{equation}
\tilde{W}(x,t)=\frac{\mu^\nu}{\Gamma(\nu)}\left(\frac{x^{\nu-1}}{t^{\alpha\nu}}\right)\,
\exp\left(-\mu\frac{x}{t^{\alpha}}\right).
\label{tW1}
\end{equation}
In the case $\nu=1$, (\ref{tW1}) reduces to (\ref{W1}) with $\mu>0$ and $\sigma=1$ as expected.
The new system can be considered as a kind of deformed system based on the parameter $\nu$.

\subsection{Examples with $\lambda\neq \alpha$}

For this case, the normalized solution, from Eq.~(\ref{W-soln}),
is 
\begin{align}\label{E4.5}
W(x,t)=\sqrt{\frac{\alpha-\lambda}{2\pi \sigma t^{2\alpha}}}\,
\mbox{exp}\Big\{-\frac{\alpha-\lambda}{2\sigma t^{2\alpha}}
\Big(x + \frac{\mu t^{\alpha}}{\alpha-\lambda}\Big)^2\Big\}\;,
\end{align}
where either $(\sigma>0$, $\lambda<\alpha)$ or $(\sigma<0$, $\lambda>\alpha)$ must
be satisfied. 

Again, for simplicity of presentation, let us take 
$\alpha=1/2$, $\lambda=\mu=0$ and $\sigma=1$.  The case is just the well-known diffusion equation, with
\begin{eqnarray}
&&\rho_1(z)=0,~~\rho_2(z)=1,\n\\
&&W(x,t)=\frac{1}{\sqrt{4\pi t}}e^{-\frac{x^2}{4t}}.
\label{diffuse}
\end{eqnarray}

A simple way to extend the class of diffusion solution Eq.~(\ref{diffuse}) is to look for 
a function $Q(z)$ such that the Gaussian-type integral 
$\int_{0}^{\infty} \!Q(z)\exp(-z^2/4)\,dz$ 
is integrable.   The simplest choice is $Q(z)=z^k$ with 
a non-negative integer $k>0$:
\begin{eqnarray}
I_k\equiv \int_0^{\infty} \!Q(z) e^{-\frac{z^2}{4}}\,dz
=  \left\{
  \begin{array}{ll}
2^n (2n-1)!!\sqrt{\pi},~~& k=2n;\\
~~ 2^{n-1} (n-1)!,~~& k=2n-1;
   \end{array}
  \right. 
  ~~~~(n=1,2,\ldots).
 \end{eqnarray}
 Here $(2n-1)!!\equiv 1\cdot 3\cdots (2n-1)$. This leads to two new processes on the positive half-line for $k=2n,~2n-1$ ($n=1,2,\ldots$), with the  PDF given by
\begin{equation}
\tilde{W}_k(x,t)=\frac{1}{I_k} t^{-\frac12}\left(\frac{x}{\sqrt{t}}\right)^k \,e^{-\frac{x^2}{4t}}.
\end{equation}

Note that for  $k$ even, i.e., $k=2n$, the extended system can be defined on the whole line, with
\begin{equation}
\tilde{W}_{2n}(x,t)=\frac{1}{ 2 I_{2n}} t^{-\frac12}\left(\frac{x}{\sqrt{t}}\right)^{2n}\,e^{-\frac{x^2}{4t}}.
\end{equation}
This reduces to the PDF in (\ref{diffuse}) for the well-known Brownian motion for $n=0$, with $I_0\equiv \sqrt{\pi}$.

%
%
\section{Equivalent systems}

It follows from Eqs.~(\ref{y-soln}) and (\ref{W-soln}) that any two sets of $\{\rho_1(z),\rho_2(z)\}$ 
and $\{\tilde{\rho}_1(z),\tilde{\rho}_2(z)\}$ that give the same $f(z)$ define equivalent FPEs as the PDFs are exactly the same.
Eqs.~(\ref{norm}) and (\ref{W-soln}) then imply that the normalized $W(x,t)$ in $x$ is obtained by normalized $y(z)$ in $z$.

This observation gives a general recipe for obtaining exactly solvable FPEs.  For a fixed $\alpha$, 
one looks for exactly integrable function $y(z)$, say  
by looking up those integrable definite
integrals in \cite{GR} such that $y(z)$ is regular 
and positive in the physical domain.  Then the function $f(z)$ is defined by
\begin{equation}
f(z)=\frac{d}{dz}\ln y(z).
\end{equation}
By selecting any function $\rho_2 (z)$ regular in the domain, the corresponding $\rho_1(z)$ is given by
\begin{equation}
\rho_1(z)\equiv \rho_2(z) f(z)+\rho_2^\prime(z)+\alpha z.
\label{rho1}
\end{equation}
Any choice 
of $\rho_2(z)$ and the corresponding $\rho_1(z)$ in Eq.~(\ref{rho1}) gives
 equivalent FPEs.

Particularly, the above recipe implies that for any solvable Fokker-Planck system with  
$\rho_1(z)$ and $\rho_2(z)$, there exist other solvable systems defined by any regular $\tilde{\rho_2}(z)$ and 
\begin{eqnarray}
\tilde{\rho}_1(z) \equiv \tilde{\rho_2}(z)\left(\frac{\rho_1(z)-\rho_2^\prime(z)-\alpha
z}{\rho_2(z)}\right) + \tilde{\rho_2}^\prime (z)+\alpha z.
\end{eqnarray}
The simplest choice of $\tilde{\rho}_2(z)$ is $\tilde{\rho}_2(z)=1$.
This means that many exactly solvable FPE's can be derived by studying exactly solvable FPEs with $\rho_2=1$. 

The above discussion gives a general recipe for finding exactly FPEs.  One simple way to apply this recipe is to take $y(z)=\phi_0(z)^2$, where  $\phi_0(z)$ 
 is the normalized ground state 
 wavefunction of some quantum mechanical system.  Thus all the ground states of exactly solvable quantal systems can be used to define
  the corresponding exactly solvable FPEs.  In this regard, we mention that the simplest examples of one-dimensional quantal systems are listed in \cite{Cooper}. 

We will not bore the reader by listing all the FPEs defined by $y(z)$ from  the ground states of known one-dimensional quantal systems as listed in \cite{Cooper}.
In what follows, we shall present some examples of FPEs which are related to quantal systems whose ground states involve the recently discovered exceptional
 orthogonal polynomials [6-19]. These can be viewed as deformed versions of some of the systems in 
 \cite{Cooper}.
The discoveries of these new kinds of polynomials, and the
 quantal systems related to them, have been among the most interesting developments in mathematical physics
in recent  years.
 Unlike the classical
orthogonal polynomials, these new polynomials have the remarkable
properties that they  start with degree $\ell=1,2\ldots,$
polynomials instead of a constant, and yet they still 
 form complete sets with respect to some
positive-definite measure.
For a recent review, see eg. Ref.~\cite{Que4}.

Below we shall only 
present examples related to the so-called single-indexed exceptional orthogonal polynomials.  Generalization to multi-indexed cases \cite{GKM4,OS3} is straightforward.




\subsection{FPEs related to deformed radial oscillator potential and 
 exceptional $X_\ell$ Laguerre polynomials}

The exceptional $X_{\ell}$ Laguerre polynomials ($\ell=1,2,3,\ldots$) are associated with
deformed radial oscillator potentials \cite{OS1,HOS}.  
These deformed radial
oscillators are iso-spectral to the ordinary radial oscillator.  

We are only interested in the ground states, which are given by
\be
\phi_{\ell,0}(z;g)=N_{\ell,0}(g) \phi_\ell(z;g) \xi_{\ell}(\eta; g+1),~~~
\phi_\ell(z;g)&\equiv & \frac{e^{-\frac{1}{2}z^2}
z^{g+\ell}}{\xi_{\ell}(\eta;g)},~~0<z<\infty. \label{phi-l}
\ee
Here  $g>0$, 
$\eta(z)\equiv  z^2$ and $\xi_\ell(\eta;g)$ is a deforming function.
 It turns out
there are two possible sets of deforming functions
$\xi_\ell(\eta;g)$, thus giving rise to two types of infinitely
many exceptional Laguerre polynomials, termed L1 and L2 type
\cite{OS1,HOS}. These $\xi_\ell$ are given by
 \be
  \xi_{\ell}(\eta;g)=
  \left\{
  \begin{array}{ll}
  L_{\ell}^{(g+\ell-\frac32)}(-\eta)&:\text{L1}\\
  L_{\ell}^{(-g-\ell-\frac12)}(\eta)&:\text{L2}
  \end{array}\right. .
  \label{xiL}
\ee
where $L^\alpha_n(\eta)$ ($n=0,1,2,\ldots$)  are the classical associated Laguerre polynomials.
The normalization constants%
are \cite{HOS}
\begin{equation}
  N_{\ell,0}(g)=
  \left\{
  \begin{array}{ll}
  \left[\frac{2 (g+\ell-\frac12)}{(g+2\ell-\frac12)\Gamma (g+\ell+\frac12)}\right]^{\frac12}&:\text{L1}\\[2pt]
 \left[\frac{ 2(g+\frac12)}{(g+\ell+\frac12)\Gamma (g+\ell+\frac12)}\right]^{\frac12}&:\text{L2}
  \end{array}
  \right. .
\end{equation}

 Note that the ground state wave functions $\phi_{\ell,0}(x;g)$ are degree $\ell$ polynomials in $\eta$, 
 instead of constants as in the ordinary cases.  For $\ell=0$, $\phi_{\ell,0}(x;g)$ reduces to the ground state of the ordinary radial oscillator.

The corresponding $f(z)$ is
\begin{equation}
f(z)=2\left[-z+\frac{g+\ell}{z}
+2z\left(\frac{\xi_\ell^\prime(\eta;g+1)}{\xi_\ell(\eta;g+1)}-\frac{\xi_\ell^\prime(\eta;g)}{\xi_\ell(\eta;g)}\right) \right],
\end{equation}
where $\xi_\ell^\prime(\eta;g)\equiv d\xi_\ell(\eta;g)/d\eta$.

Up to this point, one can consider this system as the extension, in the sense discussed in the last section, of the 
undeformed radial oscillator ($\ell=0$) modified by a choice of
\be
Q(z)=\frac{\xi_\ell(\eta;g+1)}{\xi_\ell(\eta;g)}.
\ee

Now if
 one choose $\rho_2(z)=1$, then $\rho_1(z)=f(z)+\alpha z$.
For the choice $\rho_2(z)=z$, one has $\rho_1(z)=zf(z)+1+\alpha z$. 
Yet as another choice, $\rho_2(z)=\xi_\ell(\eta;g)$, we have
\be
\rho_1(z)=\xi_\ell(\eta;g) f(z)+ 2z \xi^\prime_\ell(\eta;g)+\alpha z.
\ee
As before, here the prime means derivative with respect 
to the basic variable of the function, i.e. $\xi^\prime_\ell(\eta;g)=d\xi_\ell(\eta;g)/d\eta$.
Last but not least, one can also choose $\rho_2(z)=\xi_\ell(\eta;g+1)$.




\subsection{FPEs related to deformed P\"oschl-Teller potential and  exceptional $X_\ell$  Jacobi polynomials}

The above discussion can be extended to systems involving the other type of exceptional orthogonal polynomials, namely, 
 FPEs related to the deformed P\"oschl-Teller potential, which involved the newly discovered  exceptional $X_\ell$  Jacobi polynomials ($\ell=1,2,3,\ldots$).
All the steps are similar to those in the last subsection, so we only give the basic data in the following.

Just as the deformed radial oscillator case, here there 
are also two possible single-indexed deformations of the ordinary 
 P\"oschl-Teller potential.   With  appropriate redefinition, one can always set the domain of the basic variable to be $z\in [0,\pi/2]$.
 This domain is finite, and as mentioned at the end of Sect.~II, similarity solutions are possible only if the boundary point scale appropriately. 
 In this case, the right boundary of the Fokker-Planck system is moving according to $x(t)=\pi t^\alpha/2$.
  
  The basic data needed in the
 formula are \cite{HOS}:
\begin{align} 
& \eta(z)=\cos 2z, 
\n\\
&  \xi_{\ell}(\eta;g,h)=
  \left\{
  \begin{array}{ll}
  P_{\ell}^{(g+\ell-\frac32,-h-\ell-\frac12)}(\eta),\ \ g>h>0&:\text{J1}\\
  P_{\ell}^{(-g-\ell-\frac12,h+\ell-\frac32)}(\eta),\ \ h>g>0&:\text{J2}
  \end{array}\right.,\n\\
&\phi_{\ell,0}(z;g,h)=N_{\ell,0}(g,h)\phi_{\ell}(z;g,h) \xi_{\ell}(\eta; g+1,h+1),\quad 
\phi_{\ell}(z;g,h)=\frac{(\sin z)^{g+\ell}(\cos z)^{h+\ell}}{ \xi_{\ell}(\eta;g,h)},\n\\
 & N_0(g,h)=\left[\frac {2(g+h)\Gamma(g+h)}{\Gamma(g+\frac12)\Gamma(h+\frac12)}\right]^\frac12,\n\\
& N^2_{\ell,0}(g,h)=
  \,N^2_0(g+\ell,h+\ell)\times
  \left\{
  \begin{array}{ll}
  \frac{(h+\frac12)(g+\ell-\frac12)}{(h+\ell+\frac12)(g+2\ell-\frac12)}
  &:\text{J1}\\[4pt]
  \frac {(g+\frac12)(h+\ell-\frac12)}{(g+\ell+\frac12)(h+2\ell-\frac12)}
 &:\text{J2}
 \end{array}\right. .
\end{align} 
Here $P_n^{(\alpha,\beta)}(\eta)$ are
the classical Jacobi polynomials.

The corresponding $f(z)$ is
\begin{equation}
f(z)=2\left[(g+\ell)\cot z - (h+\ell)\tan z -2\sin 
2z \left(\frac{\xi_\ell^\prime(\eta;g+1,h+1)}{\xi_\ell(\eta;g+1,h+1)}-\frac{\xi_\ell^\prime(\eta;g,h)}{\xi_\ell(\eta;g,h)}\right) \right],
\end{equation}
where $\xi_\ell^\prime(\eta;g,h)\equiv d\xi_\ell(\eta;g,h)/d\eta$.

As in the last 
subsection, a set of  interesting choices of $\rho_2(z)$ are $\rho_2(z)=1, z, \sin z, \cos z, \xi_\ell(\eta;g,h)$ 
and $\xi_\ell(\eta;g+1,h+1)$.

\section*{Acknowledgements}
R.\,S. thanks Pei-Ming Ho for the hospitality at National Taiwan University.
C.L.H. is supported in part by the
National Science Council (NSC) of the Republic of China under
Grant NSC-102-2112-M-032-003-MY3.  R.\,S. is supported   in part by Grant-in-Aid for Scientific Research
from the Ministry of Education, Culture, Sports, Science and Technology
(MEXT) No.22540186.
We also acknowledge the support by the National Center for Theoretical Sciences -North branch (NCTSn) of R.O.C.

\end{document}